\documentclass[11pt]{article}

\usepackage[margin=1in]{geometry}
\usepackage[T1]{fontenc}
\usepackage[utf8]{inputenc}
\usepackage{lmodern}
\usepackage{microtype}
\usepackage{graphicx}
\usepackage{booktabs}
\usepackage{array}
\usepackage{multirow}
\usepackage{siunitx}
\usepackage{hyperref}
\usepackage[nameinlink,capitalize]{cleveref}
\usepackage{xcolor}
\usepackage{enumitem}

\newcommand{\rirsmall}{\texttt{rir\_small}}
\newcommand{\rirmedium}{\texttt{rir\_medium}}
\newcommand{\fzero}{\ensuremath{f_{0}}}
\newcommand{\nmels}{\ensuremath{n_{\text{mels}}}}

\usepackage{listings}
\lstdefinestyle{cli}{
  basicstyle=\ttfamily\footnotesize,
  breaklines=true,
  breakatwhitespace=true,
  columns=fullflexible,
  keepspaces=true,
  postbreak=\mbox{\(\hookrightarrow\)\space} 
}

\newcommand{\DatasetName}{BeepBank-500}
\newcommand{\DatasetURL}{https://doi.org/10.5281/zenodo.17172015}
\newcommand{\RepoURL}{https://github.com/mandip42/earcons-mini-500}

\newcommand{\NumClips}{300--500}
\newcommand{\SR}{48\,kHz}
\newcommand{\BitDepth}{16-bit}

\title{\DatasetName: A Synthetic Earcon Mini-Corpus for UI Sound and Psychoacoustics Research}
\author{Mandip Goswami\thanks{Work conducted independently; no proprietary or employer data used. Opinions are the author’s; affiliation for identification only.}\\
Principal Scientist, Amazon, Bellevue, WA, USA\\
\texttt{gomandip@amazon.com}}
\begin{document}
\maketitle

\begin{abstract}
We introduce \textbf{\DatasetName}, a compact, fully synthetic earcon/alert dataset (\NumClips{} clips) designed for rapid, rights-clean experimentation in human--computer interaction and audio machine learning. Each clip is generated from a parametric recipe controlling waveform family (sine, square, triangle, FM), fundamental frequency, duration, amplitude envelope, amplitude modulation (AM), and lightweight Schroeder-style reverberation. We use three reverberation settings: \emph{dry}, and two synthetic rooms denoted \rirsmall{} (“small”) and \rirmedium{} (“medium”) throughout the paper and in the metadata. We release mono \SR{} WAV audio (\BitDepth{}), a rich metadata table (signal/spectral features), and tiny reproducible baselines for (i) waveform-family classification and (ii) $f_0$ regression on single tones. The corpus targets tasks such as earcon classification, timbre analyses, and onset detection, with clearly stated licensing and limitations. Audio is dedicated to the public domain via CC0-1.0; code is under MIT. Data DOI: \href{\DatasetURL}{\DatasetURL}. Code: \url{\RepoURL}.
\end{abstract}
\section{Motivation and Scope}
Non-speech auditory icons and ``earcons'' are ubiquitous---from mobile notifications and wearable haptics-with-sound to automotive HMI beeps and accessibility UI cues. Researchers and practitioners frequently need \emph{small, rights-clean} corpora for prototyping classifiers, testing psychoacoustic features, or producing didactic figures. Large datasets exist for environmental audio and speech, but there is a gap for compact, controllable earcon sets emphasizing timbre variables (harmonicity, AM depth/rate, envelope) and simple room effects.

\DatasetName{} addresses this gap with three design goals: (1) \emph{tiny yet diverse} (few hundred items spanning key parameters); (2) \emph{deterministically reproducible} (scripted generation with seeds and full metadata); and (3) \emph{frictionless reuse} (CC0-1.0 audio; MIT code; Zenodo DOI). The dataset intentionally avoids industrial or proprietary sounds and makes no medical or safety-critical claims.

\paragraph{Intended uses.} Rapid experiments in UI earcon classification, timbre similarity/embedding analysis, robustness studies (reverb, AM), onset detection, and as a teaching/benchmarking resource. \paragraph{Out of scope.} Speech/music content, affective labeling, clinical applications, and complex room acoustics beyond lightweight Schroeder reverberation.

\section{Related Resources (Context Only)}
There are synthetic tone banks and earcon papers scattered across HCI and audio ML venues; however, many assets are (i) not centralized with a DOI, (ii) lack a clear license, or (iii) are much larger than necessary for didactic tasks. \DatasetName{} complements broader environmental corpora by focusing narrowly on parametric UI tones with a compact, fully scripted recipe. (We purposefully keep references minimal; this is a data note rather than a survey.)

\section{Generation Protocol}
\label{sec:protocol}
\subsection{Signal Chain}
The generation pipeline is: oscillator $\rightarrow$ (optional) amplitude modulation $\rightarrow$ ADSR envelope $\rightarrow$ RMS normalization $\rightarrow$ (optional) Schroeder-style reverb. All steps are implemented in simple Python/NumPy for transparency and speed.

\paragraph{Oscillators.} We provide \texttt{sine}, \texttt{square}, \texttt{triangle}, and two FM variants (\texttt{fm\_2to1}, \texttt{fm\_3to2}) using fixed ratios and moderate indices to induce controllable inharmonicity while keeping spectral content compact for short durations.

\paragraph{Fundamental frequency ($f_0$).} A small set of nominal centers (e.g., \SI{350}{Hz}, \SI{500}{Hz}, \SI{750}{Hz}, \SI{1000}{Hz}) is used for coverage across low to mid-high ranges typical of earcons.

\paragraph{Duration and envelopes.} Durations of \SIlist{100;250;500}{ms} coupled with three envelope presets (\texttt{adsr\_fast}, \texttt{adsr\_med}, \texttt{percussive}) modulate attack/decay and sustain level for percussive versus sustained cues.

\paragraph{Amplitude modulation (AM).} Optional sinusoidal AM with rate $r \in \{0,\,8,\,30\}\,$Hz and depth $d \in \{0.0,\,0.3,\,0.5\}$ simulates roughness/urgency cues common in alarms.

\paragraph{Chordal options.} Items may be single tones or simple triads (major or minor) to yield richer timbres without complicating the labeling scheme.

\paragraph{Reverberation.} Two lightweight Schroeder configurations emulate ``small'' ($\sim$\SI{0.3}{s}) and ``medium'' ($\sim$\SI{0.6}{s}) rooms via short comb and all-pass chains; a dry version is always available.

\paragraph{Normalization and peak handling.} Signals are RMS-normalized to a nominal target (e.g., \SI{-20}{dBFS}) with a hard cap at \SI{-1}{dBFS} to avoid clipping. LUFS may be computed for analysis (not used for normalization) if \texttt{pyloudnorm} is present.

\subsection{Parameter Grid}
\label{sec:grid}
A Cartesian product over: waveform family\,$\times$\,{$f_0$}\,$\times$\,duration\,$\times$\,envelope\,$\times$\,AM rate/depth\,$\times$\,chord type\,$\times$\,reverb kind yields a superset from which \NumClips{} items are sampled (deterministic shuffle). See \cref{tab:grid} for an overview.

\begin{table}[t]
  \centering
  \caption{Parameter grid summary.}
  \label{tab:grid}
  \begin{tabular}{@{}ll@{}}
  \toprule
  Factor & Values \\
  \midrule
  Waveform & sine, square, triangle, fm\_2to1, fm\_3to2 \\
  $f_0$ (Hz) & 350, 500, 750, 1000 \\
  Duration (ms) & 100, 250, 500 \\
  Envelope & adsr\_fast, adsr\_med, percussive \\
  AM rate (Hz) & 0, 8, 30 \\
  AM depth & 0.0, 0.3, 0.5 \\
  Chord & single, major triad, minor triad \\
  Reverb & dry, \rirsmall{}, \rirmedium{} \\
  \bottomrule
  \end{tabular}
\end{table}

\subsection{File Format and Splits}
All audio is mono, \SR{}, \BitDepth{} PCM WAV. Deterministic train/val/test splits are assigned by hashing the filename to ensure stable partitions across regenerations. 

\section{Metadata Schema and Measures}
\label{sec:metadata}
Each row in \texttt{metadata/metadata.csv} describes one clip. Columns and units are summarized in \cref{tab:metadata}. Features include simple spectral statistics and proxies for roughness/inharmonicity (explicitly flagged as proxies). LUFS is optional (blank if not computed). Re-computation scripts are included. v1.0.0 contains 400 clips, split train/val/test = 80/10/10 by deterministic filename hash.

\begin{table*}[t]
  \centering
  \caption{Metadata schema (columns in \texttt{metadata.csv}). Units embedded in names for clarity.}
  \label{tab:metadata}
  \sisetup{table-number-alignment = center}
  \begin{tabular}{@{}llp{8.2cm}@{}}
    \toprule
    Column & Type & Description \\
    \midrule
    file & string & Relative WAV path (filename). \\
    split & {train/val/test} & Deterministic partition via filename hash. \\
    sr\_hz & integer & Sample rate (\SR{}). \\
    bit\_depth & integer & PCM bit depth (\BitDepth{}). \\
    duration\_ms & integer & Duration in milliseconds. \\
    peak\_dbfs, rms\_dbfs & float & Peak and RMS levels. \\
    lufs & float/empty & Integrated loudness (if computed). \\
    waveform & categorical & sine, square, triangle, fm\_2to1, fm\_3to2. \\
    \fzero{}\_hz & integer & Nominal fundamental frequency. \\
    chord & categorical & single, major, minor. \\
    am\_rate\_hz, am\_depth & float & AM parameters. \\
    envelope & categorical & adsr\_fast, adsr\_med, percussive. \\
    reverb & categorical & dry, rir\_small, rir\_medium. \\
    spec\_centroid\_hz & float & Spectral centroid. \\
    bandwidth\_hz & float & Spectral bandwidth (stdev). \\
    zcr & float & Zero-crossing rate. \\
    inharmonicity\_proxy & 0/1 & 1 if chordal; 0 if single tone. \\
    roughness\_proxy & float & Equals AM depth as a simple proxy. \\
    attack\_ms, release\_ms & integer & Envelope edge durations (preset-dependent). \\
    seed & integer & RNG seed used for generation. \\
    version & string & Dataset semantic version (e.g., 1.0.0). \\
    \bottomrule
  \end{tabular}
\end{table*}

\section{Baselines and Example Analyses}
\label{sec:baselines}
We provide two minimal baselines intended to verify signal diversity and facilitate quick comparisons.

\paragraph {Waveform-family classification.}

Features: log-mel spectrogram (\nmels{} = 64) with global mean/variance pooling.
Model: logistic regression. \textbf{Test accuracy: 81.1\%}.

\paragraph{$f_0$ regression (single tones).}
We evaluate a parameter-free baseline (YIN + median over frames) on all single-tone items
across durations, AM settings, and reverbs (\(n=111\)). The error distribution is heavy-tailed:
\(\textbf{MAE}=63.66\,\text{Hz}\) while \(\textbf{MedAE}=0.22\,\text{Hz}\),
consistent with occasional octave/subharmonic errors under FM and reverberant/AM conditions.
To summarize robustness we additionally report the proportion within a musical tolerance
(\(\pm 1\) semitone, \(2^{1/12}-1 \approx 5.95\%\) of \(f_0\)): \textbf{80.2\%}.
See Table~\ref{tab:f0}.

\begin{table}[t]\centering
\caption{$f_0$ baseline summary on single tones.}
\label{tab:f0}
\begin{tabular}{@{}lccc@{}}
\toprule
Subset & $n$ & MAE (Hz) & MedAE (Hz) \\
\midrule
All single tones (any AM/reverb/duration) & 111 & 63.66 & 0.22 \\
\addlinespace
\multicolumn{4}{@{}l@{}}{\emph{Robustness: } \% within $\pm$1 semitone = \textbf{80.2}\%} \\
\bottomrule
\end{tabular}
\end{table}

\begin{figure}[t]
  \centering
  \includegraphics[width=0.95\linewidth]{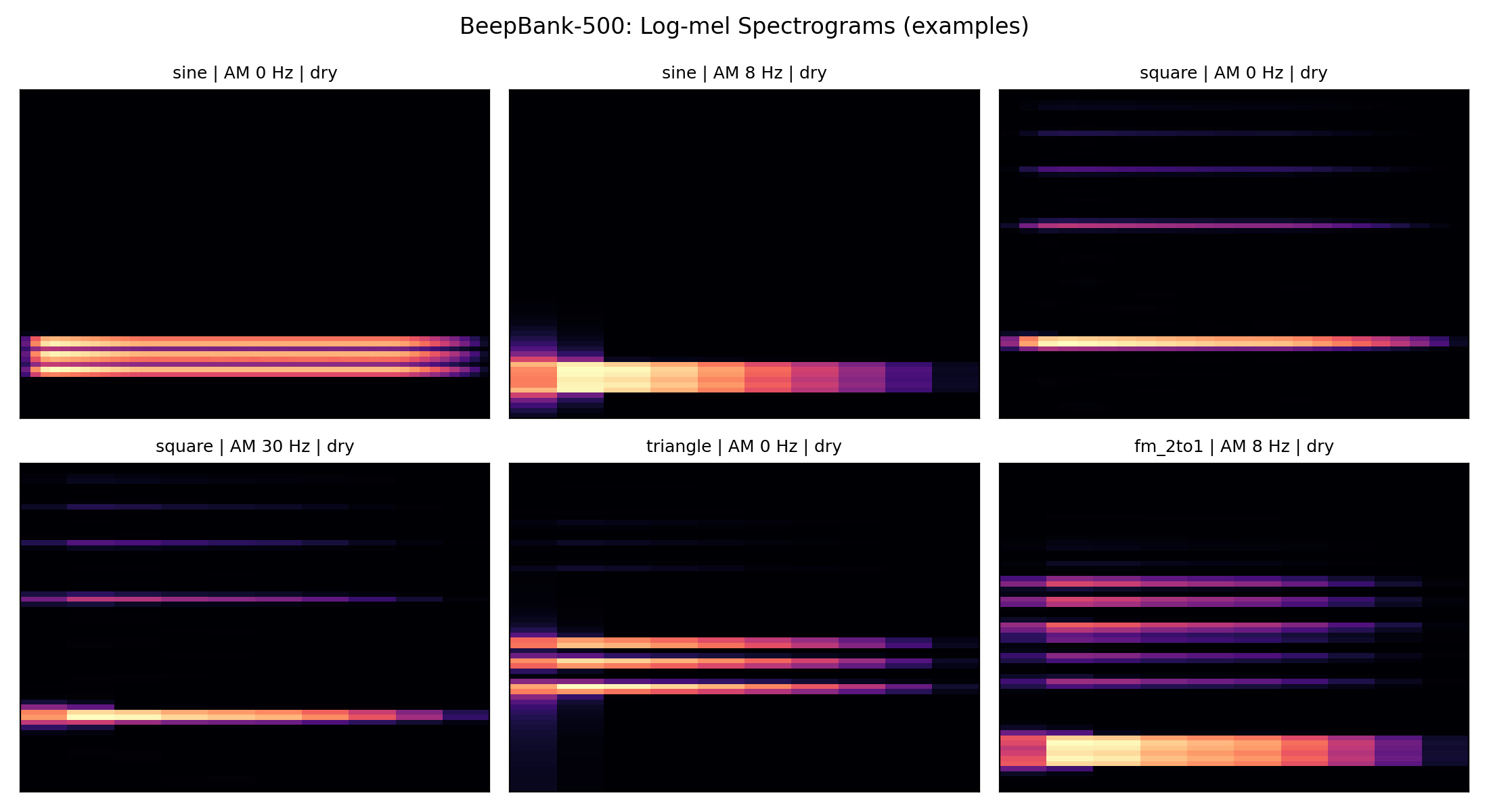}
  \caption{Example log-mel spectrograms across waveform families and AM settings.}
  \label{fig:spects}
\end{figure}

\paragraph{Reproducibility.} Scripts, seeds, and exact dependencies are provided in the repository. Baseline output JSON files capture metrics for inclusion in papers.

\section{Ethics, Licensing, and Intended Use}
\label{sec:ethics}
\textbf{Licensing.} All \emph{generated audio} is dedicated to the public domain under CC0-1.0; see \href{\DatasetURL}{Zenodo} and \texttt{metadata/LICENSES.md}. \emph{Code} is MIT-licensed. If later versions add third-party CC-BY assets, full attributions will be recorded in \texttt{LICENSES.md}.

\textbf{Intended use.} Research and education on earcon design, timbre features, simple robustness testing (e.g., to small reverbs). Not for safety-critical alerts or clinical purposes.

\textbf{Known limitations and risks.} Synthetic signals may not capture perceptual subtleties of human-designed earcons. Reverb is schematic; psychoacoustic measures are proxies unless explicitly computed. No private or sensitive information is present.

\section{Limitations and Future Work}
We intentionally prioritize compactness and controllability over ecological breadth. Future releases may add: (i) HRTF-based spatialization for 3D earcons; (ii) additional envelopes and FM indices; (iii) measured room impulse responses; (iv) optional subjective preference data (user studies); and (v) expanded $f_0$ sets and micro-variations.

\section*{Availability and Citation}
\label{sec:availability}
\textbf{Dataset (Zenodo DOI):} \url{\DatasetURL}\\
\textbf{Code:} \url{\RepoURL}

If you use \DatasetName{}, please cite the dataset DOI and this data note. A plain-text citation is provided in the repository \texttt{CITATION.cff}.

\section*{Reproducibility Checklist}
\begin{itemize}[leftmargin=*,itemsep=2pt]
  \item \textbf{Data generation code:} Provided in \texttt{code/generate\_earcons.py} (deterministic seeds).
  \item \textbf{Metadata schema:} Documented in \cref{sec:metadata} and shipped as CSV.
  \item \textbf{Baselines:} Minimal scripts with fixed preprocessing; JSON metrics are emitted.
  \item \textbf{Licenses:} CC0-1.0 for audio, MIT for code; clearly indicated in files and record.
  \item \textbf{Versioning:} Semantic version tags (e.g., v1.0.0) with CHANGELOG entries.
\end{itemize}

\appendix
\section{Quick Start (CLI)}
\label{app:quickstart}
\begin{lstlisting}[style=cli]
python -m venv .venv; source .venv/bin/activate; pip install -r requirements.txt

python code/generate_earcons.py --outdir audio --meta metadata/metadata.csv --seed 13 --target_n 400

python code/baselines/classify_waveform.py --audio_dir audio --meta metadata/metadata.csv
python code/baselines/f0_regression.py --audio_dir audio --meta metadata/metadata.csv
\end{lstlisting}

\section{Minimal BibTeX for the Dataset}
\label{app:bib}
\begin{verbatim}
@dataset{goswami_beepbank500_2025,
  author  = {Goswami, Mandip},
  title   = {BeepBank-500: A Psychoacoustic Earcon Mini-Corpus},
  year    = {2025},
  version = {1.0.0},
  doi     = {10.5281/zenodo.17172015},
  url     = {https://doi.org/10.5281/zenodo.17172015}
}
\end{verbatim}

\end{document}